\documentclass[sigconf]{acmart}

\AtBeginDocument{%
  }

\setcopyright{acmlicensed}
\copyrightyear{2018}
\acmYear{2018}
\acmDOI{XXXXXXX.XXXXXXX}
\acmConference[Conference acronym 'XX]{Make sure to enter the correct
  conference title from your rights confirmation email}{June 03--05,
  2018}{Woodstock, NY}
\acmISBN{978-1-4503-XXXX-X/2018/06}

\usepackage{tcolorbox}
\usepackage{xspace}

\acmConference[FSE '26]{Foundations of Software Engineering}{November 2026}{Your Location}
\acmYear{2026}
\acmISBN{978-1-4503-XXXX-X/26/11}
\acmDOI{10.1145/XXXXXXX.XXXXXXX}
\acmPrice{15.00}

\begin{document}

\title{When Elo Lies: Hidden Biases in Codeforces-Based Evaluation of Large Language Models}

\author{Shenyu Zheng\textsuperscript{1},Ximing Dong\textsuperscript{1}, Xiaoshuang Liu\textsuperscript{1}, Gustavo Oliva\textsuperscript{1}, Chong Chun Yong\textsuperscript{4}, Dayi Lin\textsuperscript{1}, Boyuan Chen\textsuperscript{1}, Shaowei Wang\textsuperscript{2}, Ahmed E. Hassan\textsuperscript{3} \\
\textsuperscript{1}Centre for Software Excellence, Huawei, Canada \\
\textsuperscript{2} Department of Computer Science, University of Manitoba, Canada \\
\textsuperscript{3}School of Computing, Queen's University, Canada\\
\textsuperscript{4} School of Information Technology, Monash University Malaysia, Malaysia\\
\texttt{shenyu.zheng@h-partners.com,\{ximing.dong,liuxiaoshuang4,gustavo.oliva, dayi.lin,boyuan.chen1\}@huawei.com, chong.chunyong@monash.edu, shaowei.wang@umanitoba.ca, ahmed@cs.queensu.ca}\\
}

\renewcommand{\shortauthors}{Zheng and Dong, et al.}

\begin{abstract}
As Large Language Models (LLMs) achieve breakthroughs in complex reasoning, Codeforces-based Elo ratings have emerged as a prominent metric for evaluating competitive programming capabilities. However, these ratings are often reported without critical experimental details, leading to significant discrepancies—illustrated by recent reports where the same model version’s score fluctuated by nearly 500 points.
This paper presents a systematic empirical study on the hidden factors biasing Elo evaluations: (1) the temporal ordering of submissions, (2) contest difficulty selection, and (3) run-to-run stochastic variability of LLMs. Utilizing a controlled benchmark of 37 recent Codeforces contests and 13,691 generated test cases, we demonstrate that Elo scores are highly sensitive to these parameters. Our findings reveal that varying submission orders can shift scores by 394 points, while contest selection can cause differences of up to 1,122 points for the same model. Run-to-run performance exhibits substantial instability, with a maximum difference of 349 points in mean scores observed when evaluating identical contests. We conclude that direct Elo comparisons are unreliable and potentially misleading without strict standardization and transparent reporting of experimental settings. 

\end{abstract}











\newcommand{\ie}{\emph{i.e.,}\xspace}
\newcommand{\eg}{\emph{e.g.,}\xspace}

\newcommand{\sw}[1]{\textcolor{red}{{\it [Shaowei says: #1]}}}

\newcommand{\xm}[1]{\textcolor{green}{{\it [Ximing says: #1]}}}

\newcommand{\sy}[1]{\textcolor{blue}{{\it [Shenyu says: #1]}}}

\newcommand{\ourtool}{\textbf{SemanticSpec}\xspace}

\newcommand*\circled[1]{\tikz[baseline=(char.base)]{
            \node[shape=circle,fill,inner sep=1pt,font=\footnotesize] (char) {\textcolor{white}{#1}};}}

\maketitle


\section{Introduction}\label{sec:intro}

With the increasing capabilities of existing LLMs and breakthroughs in reasoning models like Deepseek R1, OpenAI o1~\cite{jaech2024openai} and o3~\cite{openai2025o3}, there is a growing need to develop benchmarks that effectively test their sophisticated reasoning abilities. Math and coding are two evaluation methods for this purpose, as they provide accurate and easily verifiable feedback. Recent work has begun to leverage Codeforces contests to benchmark large language models (LLMs) on competition-level code generation.

However, existing Codeforces-based evaluations often omit critical experimental details that can substantially influence the resulting Elo scores. In particular, key factors such as the specific selection of contest problems, the temporal order of failed and successful submissions, and the stochastic variability across different runs of the same LLM are frequently under-specified or entirely undocumented. This lack of transparency fundamentally undermines the reliability and reproducibility of cross-model comparisons. The consequences of these omissions can be severe. For example, in the initial release of DeepSeek-R1~\cite{deepseek_r1_0120}, the model was reported to achieve an Elo score of 2,029, whereas a later report~\cite{deepseek_r1_0528} revised the score for the same model version down to 1,530 - \textbf{a discrepancy of nearly 500 Elo points}. 

Designed to mirror competitive programming environments, the Elo system updates a participant’s rating based on multiple interacting factors. Beyond the number of problems solved, it penalizes unsuccessful submissions that precede a successful one, with each failed attempt incurring a rating deduction that partially offsets the eventual gain~\cite{djm03178_2024_scoring}. Moreover, problem difficulty is explicitly incorporated into the rating update, meaning that the choice of contests and divisions directly affects the final Elo score. These implicit variables introduce nontrivial noise and bias, complicating the interpretation of Elo as a stable measure of a model’s competitive-level reasoning ability.
The problem is further exacerbated by the well-known instability of LLM outputs across runs~\cite{holtzman2019curious,wang2022self,dodge2022measuring}. Even using the same temperature and decoding trajectories can lead to markedly different solution attempts, numbers of failures, and even sets of solved problems.
\textbf{Such a variance (i.e., 500 Elo points) is difficult to attribute to genuine model improvements or regressions alone and instead points to substantial sensitivity to evaluation settings.}
Consequently, unlike standard static benchmarks, directly comparing Elo scores across models—or even across different evaluations of the same model—without carefully controlling and reporting experimental settings can be misleading and potentially unfair.

In this paper, we present an empirical study that systematically examines hidden factors that bias Elo-based evaluations of LLMs on Codeforces. We focus on three sources of variation: (1) the temporal ordering of failed and accepted submissions, (2) the selection of contests with different difficulty levels (e.g., divisions), and (3) run-to-run variability of the same LLM. To enable controlled analysis, we construct a benchmark comprising 37 Codeforces contests collected between February and September 2025 and generate 13,691 LLM-produced test cases to support local and consistent correctness verification. Our experiments on five advanced models (e.g., 04-mini and DeepSeek v3.1) demonstrate that Elo scores are highly sensitive to experimental configurations. For instance, varying submission orders can result in a maximum 394-point difference on the same model and contest. Selecting different divisions of contests can lead to a maximum of 1,122-point difference when evaluating on the same model. Run-to-run performance exhibits substantial instability, with a maximum difference of 349 points in mean scores observed when evaluating identical contests. In one single contest, we observe a 1,348-point difference from two runs by DeepSeek v3.1. Such variance is large enough to reorder model rankings, meaning Elo-based leaderboards can change substantially under different (often unreported) evaluation settings. Consequently, we argue that direct Elo comparisons are unreliable and potentially misleading unless the underlying experimental 
settings are strictly standardized and reported. The data and code that support the findings of this study will be made public upon publication. 

\section{Background}\label{sec:background}


The Elo rating system is a method for calculating the relative skill levels of players in zero-sum games (traditionally chess, but now widely used in video games, football, and even ranking AI models). The Codeforces rating system adapts the traditional 1-on-1 Elo system (originally built for chess) to work in a ``battle royale'' format that allows hundreds of people to compete simultaneously~\cite{Mirzayanov2015Rating}.
Every community member is assigned an Elo score, $r_i$, which serves as a numerical representation of their skill. The goal of the rating system is to ensure that a player's score accurately predicts their probability of success against others. The winning probability ($P_{i,j}$) between two participants, $i$ and $j$, is defined by the probability that participant $i$ will outperform participant $j$ based on following equation:
$$P_{i,j} = \frac{1}{1 + 10^{\frac{r_j - r_i}{400}}}$$

The next question is how we update the Elo score after knowing the performance of each participant after a contest? Specifically, let us assume there are $n$ participants in a contest with ratings $r_i$ for $i = 1, 2, ..., n$. For ease of mathematical representation, assume participants are ranked from best to worst in terms of performance in this contest. Position the model’s performance within this ranking, denoting the model’s rank as $m$ (where $1 \leq m \leq n + 1$). Suppose the model’s expected Elo score is $r$, according to the definition of the Elo rating~\cite{elo1978rating}, we have the following equation by following previous study~\cite{quan2025codeelo}:

$$m = \sum_{i=1}^{n} \frac{1}{1 + 10^{(r -r_i)/400}}$$

Since the expected rank function ($m$) is monotonic (meaning as your score goes up, your expected rank numerically goes down and gets better). Now we can use binary search to search the expected score $r$ for all participants based on their ranking $m$ by following the previous study~\cite{quan2025codeelo}. For instance, suppose a participant's ranking is 150th. If we guess an Elo score of 1,500, according to the equation, we get a ranking of 500th, which is too low. So you should search for a higher score $\frac{(1500+ max\_rating)}{2}$. 


Codeforces utilizes a specific scoring system where a participant's rank is determined by their final score, which is calculated based on the number of solved problems and the accumulated penalty~\cite{djm03178_2024_scoring}. This penalty is influenced by two significant factors: 1) the submission time, defined as the number of minutes elapsed from the start of the contest until a problem is solved, and 2) the number of incorrect submissions made prior to the first ``Accepted'' submission for that problem. Therefore, even if two participants solve the same number of problems, the one who solves them faster and with fewer incorrect attempts will incur a lower penalty, resulting in a higher overall rank.



\section{Experiment Design}\label{sec:method}

In this study, we aim to investigate the impact of three factors on Elo scores by answering the following three RQs:
\begin{itemize}
    \item \textbf{RQ1:} What is the impact of the temporal ordering of failed and accepted submissions on Elo scores? 
    \item \textbf{RQ2:} What is the impact of selecting contests of different difficulty levels (i.e., divisions) on Elo scores?
    \item \textbf{RQ3:} What is the impact of run-to-run variability of an LLM on Elo Scores?
\end{itemize}

The methodology of our study mainly has two stages: 1) benchmark construction, and 2) local model evaluation. 

\subsection{Benchmark Construction}

\begin{figure*}
    \centering
    \includegraphics[width=\linewidth]{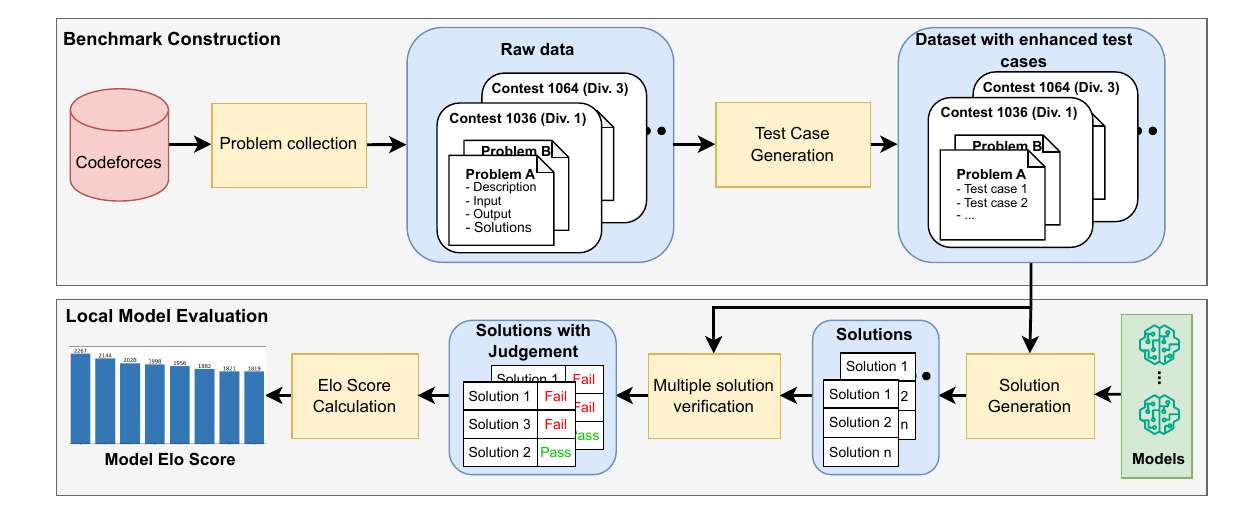}
    \vspace{-0.3in}
    \caption{Overview of our experimental methodology}
    \label{fig:framework}
    \vspace{-0.2in}
\end{figure*}

Existing official Codeforces platforms and Codeforces-based benchmarks~\cite{quan2025codeelo} have key limitations: \circled{1} \textbf{Limited access for evaluation}. LLM-generated solutions need to be judged by Codeforces through its API, preventing end users from large-scale reproducibility due to API limits. \circled{2} \textbf{Hidden test cases}. Released datasets typically provide only problem statements and metadata (including a limited number of sample test cases), while official test cases remain inaccessible, making local reproduction and systematic evaluation of alternative models impossible.
To enable our study, we introduce a pipeline that constructs the benchmark as shown in Figure~\ref{fig:framework}. We first collect problems and their partial solutions from Codeforces. Next, we generate test cases for each collected program by leveraging LLMs. 
Our pipeline allows practitioners to evaluate the models locally, with better flexibility. 

\subsubsection{Problem Collection}

Our contest problems are sourced from the official Codeforces platform\footnote{\url{https://codeforces.com}}
. The scraped problems, originally in raw HTML, were parsed into sections such as the problem description, input/output format, examples, notes, and algorithm tags. We also collect correct solutions from contest blog posts, which provide official answers. For instance, the solutions for Codeforces Round 1005 (Div. 2)\footnote{\url{https://codeforces.com/blog/entry/138912}}
. Finally, to simulate the Codeforces Elo system, we gather the problem resolution status for each participant, which allows comparison when evaluating LLMs.

\subsubsection{Test Case Collection and Generation}
While we collect sample test cases provided on Codeforces problem page, those test cases are primarily intended to illustrate problem statements and input formats, and thus cover only a limited subset of possible execution paths. As a result, solutions that pass the sample tests may still fail on corner cases or extreme inputs that exploit boundary conditions, uncommon configurations, or full input constraints. Therefore, we leverage LLMs' powerful coding ability to generate test inputs~\cite{chen2024chatunitest,ryan2024code}. The proposed prompt is designed to systematically address the above-mentioned limitation by instructing the LLM to generate difficult test inputs that go beyond typical examples. By explicitly emphasizing tricky edge cases and maximal use of input limits, the prompt encourages the generation of adversarial and stress-inducing inputs that are more likely to expose logical errors, incorrect assumptions, or performance bottlenecks in candidate solutions. 

\begin{tcolorbox}[
    title=Prompt for Test Case Generation,
    colback=gray!5,
    colframe=black,
    fonttitle=\bfseries,
    sharp corners,
    boxrule=0.4pt
]
\footnotesize
\ttfamily
Generate \{idea\_test\_count\} difficult inputs for the following Codeforces
problem to evaluate the correctness of the solution. Each input should be a string with a format similar to the provided examples.
The generated inputs should focus on tricky edge cases and make full use of the
input constraints to thoroughly stress-test the solution.

Return the results directly in JSON, wrapped in Markdown as follows:

\medskip
\texttt{```json} \\
\texttt{["input\_a", "input\_b", ...]} \\
\texttt{```}
\medskip

\textbf{Problem:}\\
\textit{\{problem\}}
\end{tcolorbox}

Note that we only use LLMs for generate test inputs, we use the official correct solution to obtain correct output to form the oracle.


\subsection{Local Model Evaluation}

\subsubsection{Multiple Solutions Verification}

While most competitive programming problems require a unique output, a significant proportion allow for multiple valid solutions (e.g., differing element orders or formats). For example, the Max Sum OR problem admits various valid arrays for a single input pair~\footnote{\url{https://codeforces.com/problemset/problem/2146/D1}}. In these instances, simple string comparison against a reference is insufficient, necessitating a solution verifier to programmatically validate results.

To identify such problems, we analyze accepted submissions from Codeforces status pages. We collect three distinct accepted solutions per problem and compare their outputs for identical test cases. Discrepancies between these outputs signal that the problem allows for multiple valid solutions.

Manually authoring verification scripts is impractical due to the highly problem-specific logic required. Instead, we leverage DeepSeek-R1-250528 to automatically generate these verifiers. By conditioning the model on the problem description and a set of verified input-output pairs, we produce a programmatic script capable of validating candidate solutions. We opt for script-based verification rather than direct LLM judging to mitigate the risk of hallucinations and logical inconsistencies inherent in generative models. This automated approach ensures a stable, reliable, and scalable evaluation framework across our benchmark.

\subsubsection{Elo Score Calculation}

We calculate the Elo score for each model following the approach introduced in Section~\ref{sec:background}. Note that, to simulate the real-world scenario, we collected the participants for each problem and used them as the competitors when calculating Elo scores for models.

\subsection{Statistics of Benchmark}

Following our methodology, we first collected a dataset of 79 contests held between February and September 2025, comprising 571 individual problems. Of these, 536 included official solutions from editorial blogs. To ensure the reliability of problems with multiple valid solutions, we employed an LLM to generate solution verifiers. Through this process, 111 problems were excluded as unverifiable, leaving a final set of 37 complete contests and 260 problems. For those 260 problems, we initially extracted 649 sample test cases directly from the Codeforces problem pages. To increase test coverage, we leveraged LLMs to augment the test suite, resulting in a total of 13,691 test cases. The dataset includes 3 Div. 1, 21 Div. 2, 8 Div. 3, and 3 Div. 4 contests.

\subsection{Base LLMs}
We evaluate a set of recent advanced models representing both open-source and commercial systems, including DeepSeek v3.1, Qwen3 235b A22b, MiniMax-M2, Kimi K2, and o4 Mini. For all models, we use their thinking capability.

\subsection{Approaches of RQs}

\subsubsection{Approach of RQ1}


Unlike order-independent $Pass@n$, Elo evaluation penalizes failed attempts preceding a success, making it sensitive to submission sequences, which is a factor largely overlooked in existing literature. To quantify this, we compare two cases: the \textbf{optimal-case} (success submitted first) and the \textbf{worst-case} (all failures submitted first). We select eight contests across four divisions (two per division) from our constructed dataset and evaluate how this order sensitivity affects the Elo scores of the models. More specifically, we examine the delta scores between optimal and worst cases on the same contest. We also investigate how the impact changes as the number of generated solutions ($n$) increases in one run. We test three values (i.e., 3, 6, and 9) for $n$.

\subsubsection{Approach of RQ2}
As discussed in Section~\ref{sec:background}, Codeforces-style Elo updates depend on final standings (rank vs expected rank). Contest division (Div.1–Div.4) correlates with both the difficulty distribution of the problem set and the strength/composition of the participant pool, which can systematically shift standings and thus Elo outcomes. Unlike standard benchmarks that treat all problems equally, solving a more difficult problem yields a greater rating gain. To quantify the effect of problem difficulty, we select two contests from each division and evaluate all studied models on them. We then compare each model’s Elo scores across divisions. We fix the number of generated solutions to $n=3$ and adopt the worst-case submission order.

\subsubsection{Approach of RQ3}
The instability of LLM outputs across runs is a well-known problem~\cite{holtzman2019curious,wang2022self,dodge2022measuring}. Even using the same temperature and decoding trajectories can lead to markedly different solution attempts, numbers of failures, and even sets of solved problems. To quantify the effect of run-to-run variation, we select two contests from each division and evaluate all studied models on them. We run the models three times and compare the results across runs. We fix the number of generated solutions to $n=3$ and adopt the worst-case submission order.

\section{Results}\label{sec:results}

\begin{figure*}[h!]
    \centering
    \includegraphics[width=\linewidth]{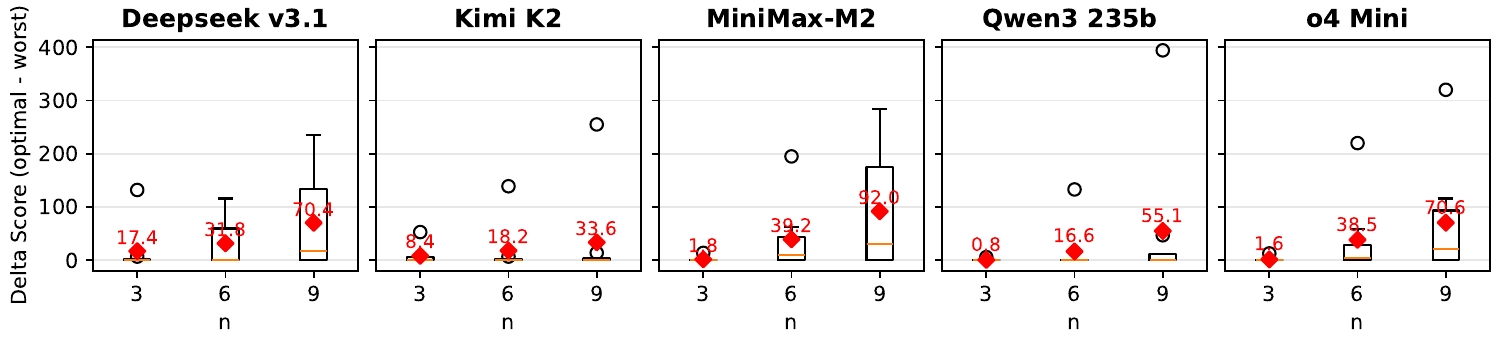}
    \vspace{-0.3in}
    \caption{Delta Elo scores (i.e., optimal-case - worst-case) of evaluated contests for different models along with different numbers of output solutions ($n$). Mean value is marked as a diamond.}
    \label{fig:elo_deltas}
\end{figure*}

\subsection{RQ1: Impact of Submission Ordering}\label{sec:Order}


\textbf{Varying submission orders resulted in a maximum Elo delta of 394 points for the same model.} As shown in Figure~\ref{fig:elo_deltas}, substantial Elo gaps exist between optimal and worst-case scenarios across all models. For example, Qwen3 235b exhibits a 394-point difference on CF Round 1028 (Div. 2). Other models show similar sensitivity, with deltas of 236 (DeepSeek v3.1), 284 (MiniMax-M2), and 320 (o4 Mini).

\textbf{The submission order bias correlates positively with the number of LLM-generated outputs.} For DeepSeek v3.1, the average Elo delta grows from 17.4 to 70.4 as $n$ increases, with a corresponding rise in standard deviation from 46.3 to 93.3. This occurs because a higher $n$ increases the likelihood of failures occurring before a correct solution is submitted. Consequently, submission order bias is not only significant but also amplifies with $n$.

\subsection{RQ2: Impact of Difficulty Level}\label{sec:Devision}

\begin{figure*}
    \centering
    \includegraphics[width=1\linewidth]{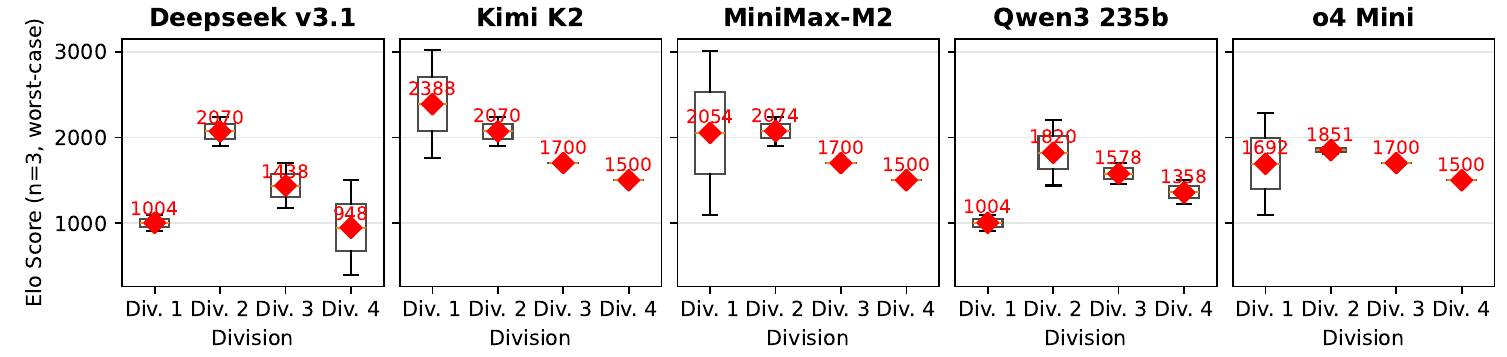}
     \vspace{-0.3in}
    \caption{Elo scores across four divisions for evaluated models. Mean value is marked as a diamond.}
    \label{fig:divs}
\end{figure*}

\textbf{Varying divisions can result in large Elo score variation, up to a maximum of 1,122-point.}
Figure~\ref{fig:divs} presents the boxplots of Elo scores for each evaluated model on different divisions.
Even for identical models facing different difficulty tiers, Elo scores vary dramatically. For instance, Deepseek v3.1 scores 2070 in Div.2, but only 948 in Division 4, which leads to a 1,122-point difference within a single model. MiniMax-M2 exhibits similar variation: 2054 in Division 1 versus 1500.5 in Division 4 (554-point difference). Kimi K2 ranges from 2388.5 (Division 1) to 1500.5 (Division 4). These substantial disparities demonstrate that problem difficulty profoundly impacts individual model performance.

\textbf{Even identical division could result in remarkably different Elo scores on the same model.} For instance, in Division 1, Deepseek v3.1 scores 1098 on CF Round 1012 but 909 on CF Round 1028—a 189-point gap for the same difficulty tier. Division 2 shows similar inconsistency: Deepseek v3.1 achieves 2242 on Edu CF Round 174 but only 1898 on CF Round 1019. MiniMax-M2 exhibits even larger within-division variation (standard deviation of 1351.99 in Division 1), and this pattern repeats across all models. Kimi K2, Qwen3 235b, and o4 Mini all demonstrate substantial score fluctuations within individual divisions, typically in difficult divisions, such as Div. 1.




\subsection{RQ3: Impact of Run-to-Run Variability of LLMs}\label{sec:run}

\begin{figure*}
    \centering
    \includegraphics[width=\linewidth]{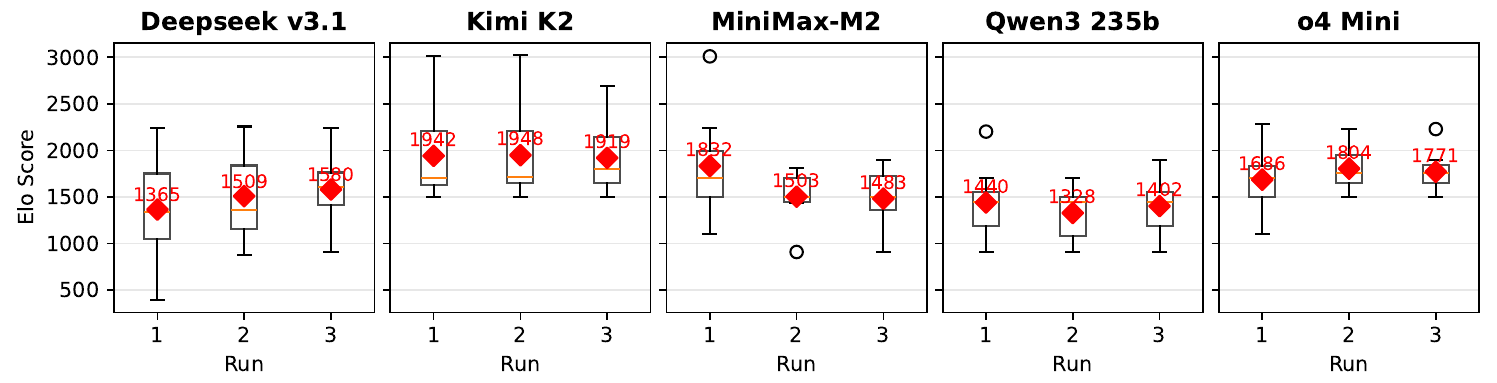}
     \vspace{-0.3in}
    \caption{Elo scores of evaluated contests across three runs for evaluated models. Mean value is marked as a diamond.}
    \label{fig:run2run}
    \vspace{-0.15in}
\end{figure*}

\textbf{LLMs' Run-to-run performance exhibits substantial variation when evaluating the same contest problems, with a maximum of 349 score differences on average.} Figure~\ref{fig:run2run} presents the results of Elo scores on the eight contests for each LLM across different runs. For instance, MiniMax-M2 demonstrates the most dramatic fluctuation: achieving a mean score of 1832.1 in run 1, while dropping to 1503.1 in run 2, and stabilizing at 1483.1 in run 3, while results a 349-point range across three runs on identical problems. Similarly, Deepseek v3.1 improves progressively from 
1364.6 (run 1) to 1580.0 (run 3), representing a 215-point difference. Kimi K2 achieves the most stable performance, between 1919.3 and 1948.3. 

\textbf{These score variations stem directly from models solving different subsets of problems across runs.} A model might successfully resolve a particular algorithmic problem in run 1, but fail on the same problem in run 2, despite identical problem presentations. For instance, DeepSeek V3.1 resolved one problem in a contest in Div. 1 (most challenging division) in run 1, while resolving none of the problems in the contest, while leads to a 1,348-point difference. This non-determinism indicates that model outputs are stochastic, which leads to a great the Elo score variation. 
This analysis underscores that LLM performance on competitive programming tasks is inherently 
non-deterministic. Evaluating models on a single run produces misleading conclusions. 

\section{Discussion}



\noindent\textbf{Learned Lessons.}
\textbf{ The Fragility of Elo Benchmarking} Our findings demonstrate that Elo scores are highly sensitive to experimental configurations. Specifically, the temporal ordering of submissions, the difficulty levels of selected contests, and the inherent stochastic nature of LLMs can lead to drastically different Elo scores for the same model. Consequently, we argue that direct Elo comparisons are unreliable and potentially misleading unless the underlying experimental settings are strictly standardized and reported.

We strongly advise against comparing Elo scores in isolation. For researchers seeking to use Elo as a metric, we recommend adopting a rigorous evaluation protocol: (1) utilize a standardized pipeline to fix submission ordering and contest selection, (2) explicitly identify and mitigate potential confounders, and (3) perform multiple experimental runs to account for run-to-run variability. Only through such controlled methodologies can Elo serve as a meaningful proxy for the model's competitive programming capability.

\noindent\textbf{Threats to Validity}
The evaluation of each solution relies on the test cases we generated. Incomplete coverage of niche corner cases could yield false positives, erroneously inflating Elo scores. We mitigated this by using high-quality reasoning models for generation, though we acknowledge that a gap may still exist between our local verification and the official online judge.

\section{Conclusion}\label{sec:conclusion}
This study demonstrates that LLM evaluation using Codeforces-based Elo ratings, while popular, is highly sensitive to hidden experimental variables. Our empirical analysis reveals that submission ordering, contest difficulty selection, and stochastic variability can swing a model's score by over 1,100 points, which suggests that direct comparisons without standardized reporting are meaningless. To ensure fair and reliable evaluation of reasoning models, we urge the research community to adopt standardized pipelines and transparently document all experimental parameters. 

\bibliographystyle{ACM-Reference-Format}
\bibliography{custom}

@article{quan2025codeelo,
  title={Codeelo: Benchmarking competition-level code generation of llms with human-comparable elo ratings},
  author={Quan, Shanghaoran and Yang, Jiaxi and Yu, Bowen and Zheng, Bo and Liu, Dayiheng and Yang, An and Ren, Xuancheng and Gao, Bofei and Miao, Yibo and Feng, Yunlong and others},
  journal={arXiv preprint arXiv:2501.01257},
  year={2025}
}

@article{elo1978rating,
  title={The rating of chessplayers: Past and present},
  author={Elo, Arpad E and Sloan, Sam},
  journal={(No Title)},
  year={1978}
}

@misc{djm03178_2024_scoring,
  author       = {djm03178},
  title        = {{Everything about Codeforces scoring system}},
  howpublished = {\url{https://codeforces.com/blog/entry/133094}},
  year         = {2024},
  month        = aug,
  note         = {Codeforces Blog}
}

@misc{deepseek_r1_0120,
  author = {DeepSeek-AI},
  title = {DeepSeek-R1: Incentivizing Reasoning Capability in LLMs via Reinforcement Learning},
  year = {2025},
  publisher = {Hugging Face},
  journal = {Hugging Face Model Card},
  howpublished = {\url{https://huggingface.co/deepseek-ai/DeepSeek-R1}}
}

@misc{deepseek_r1_0528,
  author = {DeepSeek-AI},
  title = {DeepSeek-R1-0528: Updates on Reasoning and Codeforces Benchmarking},
  year = {2025},
  publisher = {Hugging Face},
  journal = {Hugging Face Model Card},
  howpublished = {\url{https://huggingface.co/deepseek-ai/DeepSeek-R1-0528}}
}

@misc{Mirzayanov2015Rating,
  author       = {Mikhail Mirzayanov},
  title        = {{Open Codeforces Rating System [updated on October 2015]}},
  howpublished = {\url{https://codeforces.com/blog/entry/20762}},
  year         = {2015},
  month        = oct,
  note         = {Codeforces Blog}
}

@article{holtzman2019curious,
  title={The curious case of neural text degeneration},
  author={Holtzman, Ari and Buys, Jan and Du, Li and Forbes, Maxwell and Choi, Yejin},
  journal={arXiv preprint arXiv:1904.09751},
  year={2019}
}

@article{wang2022self,
  title={Self-consistency improves chain of thought reasoning in language models},
  author={Wang, Xuezhi and Wei, Jason and Schuurmans, Dale and Le, Quoc and Chi, Ed and Narang, Sharan and Chowdhery, Aakanksha and Zhou, Denny},
  journal={arXiv preprint arXiv:2203.11171},
  year={2022}
}

@inproceedings{dodge2022measuring,
  title={Measuring the carbon intensity of ai in cloud instances},
  author={Dodge, Jesse and Prewitt, Taylor and Tachet des Combes, Remi and Odmark, Erika and Schwartz, Roy and Strubell, Emma and Luccioni, Alexandra Sasha and Smith, Noah A and DeCario, Nicole and Buchanan, Will},
  booktitle={Proceedings of the 2022 ACM conference on fairness, accountability, and transparency},
  pages={1877--1894},
  year={2022}
}

@inproceedings{chen2024chatunitest,
  title={Chatunitest: A framework for llm-based test generation},
  author={Chen, Yinghao and Hu, Zehao and Zhi, Chen and Han, Junxiao and Deng, Shuiguang and Yin, Jianwei},
  booktitle={Companion Proceedings of the 32nd ACM International Conference on the Foundations of Software Engineering},
  pages={572--576},
  year={2024}
}

@article{ryan2024code,
  title={Code-aware prompting: A study of coverage-guided test generation in regression setting using llm},
  author={Ryan, Gabriel and Jain, Siddhartha and Shang, Mingyue and Wang, Shiqi and Ma, Xiaofei and Ramanathan, Murali Krishna and Ray, Baishakhi},
  journal={Proceedings of the ACM on Software Engineering},
  volume={1},
  number={FSE},
  pages={951--971},
  year={2024},
  publisher={ACM New York, NY, USA}
}

@article{jaech2024openai,
  title={Openai o1 system card},
  author={Jaech, Aaron and Kalai, Adam and Lerer, Adam and Richardson, Adam and El-Kishky, Ahmed and Low, Aiden and Helyar, Alec and Madry, Aleksander and Beutel, Alex and Carney, Alex and others},
  journal={arXiv preprint arXiv:2412.16720},
  year={2024}
}

@techreport{openai2025o3,
  title={OpenAI o3 and o4-mini System Card},
  author={OpenAI},
  institution={OpenAI},
  year={2025},
  month={April},
  url={https://openai.com/index/o3-o4-mini-system-card/}
}

\end{document}